\documentclass[a4paper,11pt]{article}
\pdfoutput=1

\usepackage{jheppub}

\usepackage[T1]{fontenc}

\RequirePackage{lineno}
\usepackage{epstopdf}
\newcommand{\too}{\rightarrow}
\newcommand{\EE}{e^+e^-}

\title{\boldmath Search for the decay $h_c\rightarrow\pi^0J/\psi$}

\collaborationImg{\includegraphics[width=.12\textwidth,origin=c,angle=90]{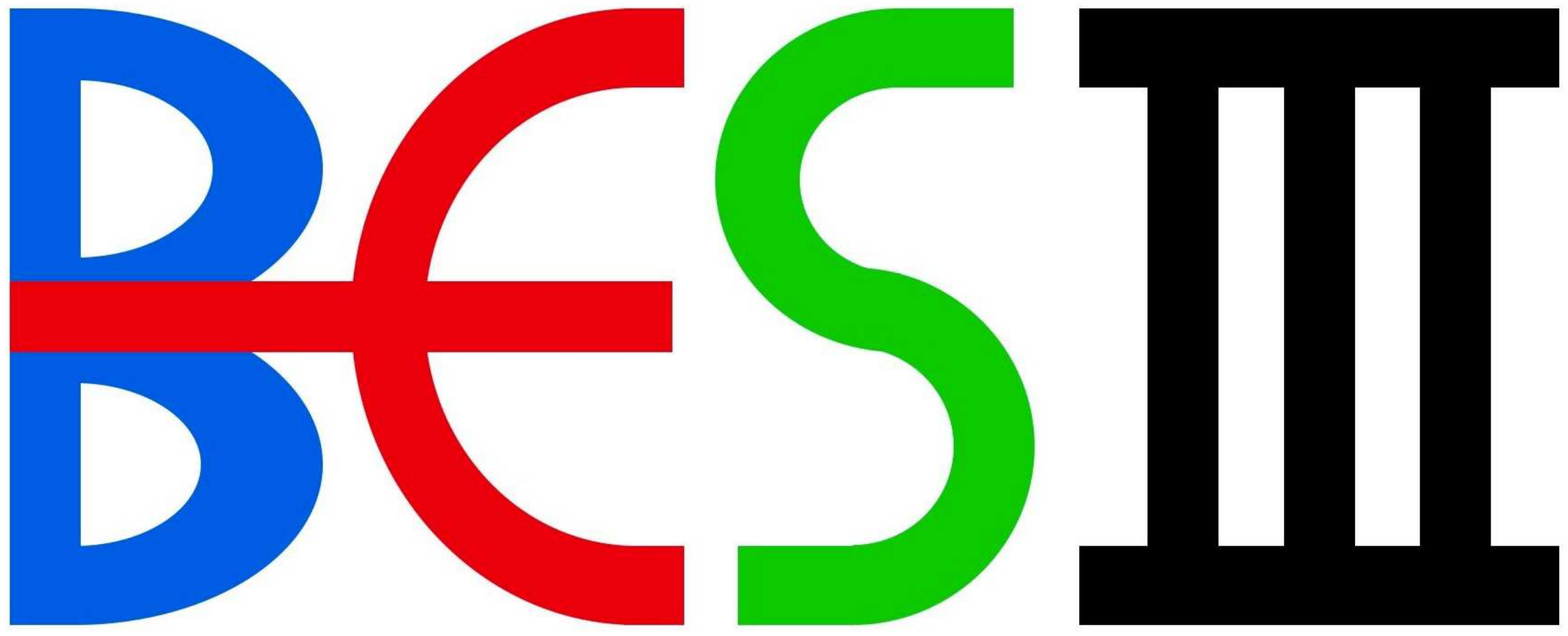}}

\collaboration{The BESIII collaboration}

\keywords{Charmonium, Branching fraction, BESIII}

\emailAdd{besiii-publications@ihep.ac.cn}

\arxivnumber{2111.13915}

\abstract{
A search for the decay $h_c\rightarrow\pi^0J/\psi$ is performed
using a sample of $h_c$ produced in the reaction $e^+e^-\rightarrow\pi^+\pi^-h_c$.
The data samples were collected with the BESIII detector at center-of-mass energies between 4.189 and 4.437 GeV, corresponding to a total integrated luminosity of 11 fb$^{-1}$.
No significant signal is observed. Upper limits on the branching ratio $\mathcal{B}(h_c\rightarrow\pi^0J/\psi)/\mathcal{B}(h_c\rightarrow\gamma\eta_c\rightarrow\gamma K^+K^-\pi^0)$ and on the branching fraction $\mathcal{B}(h_c\rightarrow\pi^0J/\psi)$ are determined to be $7.5\times10^{-2}$ and $4.7\times10^{-4}$ at $90\%$ confidence level, respectively. The latter is derived from the former using the measured branching fraction of the normalization channel. This is the first determination of the upper limit of the decay $h_c\rightarrow\pi^0J/\psi$.}

\begin{document}
\maketitle
\flushbottom


\section{Introduction}
Charmonium, the bound state of a charm quark and anticharm quark ($c\bar{c}$), plays an important role in our understanding of quantum chromodynamics (QCD), which is the fundamental theory of the strong interaction. Low energy QCD remains a field of high interest both experimentally and theoretically. All charmonium states below the open-charm ($D\bar{D}$) threshold have been observed experimentally and can be well described by QCD inspired models~\cite{quarkmodel}. However, knowledge is still sparse on the $P$-wave spin-singlet state, $h_c(1P)$~\cite{pdg}. So far, only a few decay modes of the $h_c$ have been observed, such as $h_c\too\gamma\eta_c(\eta')$~\cite{pdg}. Searches for new decay modes of the $h_c$ can provide useful information to constrain theoretical models in the charmonium region.

In 1992, the E760 Collaboration reported an evidence of the $h_c$ in the $\pi^0J/\psi$ decay mode~\cite{e760}. The decay $h_c\too\pi^0J/\psi$ was not confirmed by the successor experiment E835 with higher statistics in 2005~\cite{e835}, but E835 Collaboration found the evidence for $h_c$ in another decay mode $\gamma\eta_c$. More measurements are thus needed for clarification. In addition, several theoretical articles have addressed the decay $h_c\too\pi^0J/\psi$, with predictions for the partial width around several keV~\cite{theory1, theory2, theory3, theory4}. Until now, there is no experimental result to confirm this.

Many studies of the $h_c$ have been performed using the reaction $\psi(3686)\rightarrow\pi^0h_c$ from the $\psi(3686)$ data sample at BESIII~\cite{hc1, hc2, hc3}. However, it is hard to search for the decay $h_c\too\pi^0J/\psi$ using the $\psi(3686)$ data sample due to the large background from $\psi(3686)\too\pi^0\pi^0J/\psi$. BESIII has observed a sizeable cross section for the process $\EE\too\pi^+\pi^-h_c$ between 4.189 and 4.437 GeV~\cite{pipihc}. This process is advantageous in the search for the decay $h_c\too\pi^0J/\psi$, because it avoids the dominant background of $\psi(3686)\too\pi^0\pi^0J/\psi$ present in the $\psi(3686)$ decay.

In this paper, a search for $h_c\too\pi^0J/\psi$ is reported using $e^+e^-\rightarrow\pi^+\pi^-h_c$ events from
data samples collected at center-of-mass energies between $4.189$ and $4.437$ GeV with the BESIII detector~\cite{besiii} with a total integrated luminosity of 11 fb$^{-1}$.
The $J/\psi$ is reconstructed in its decay to a $\ell^+\ell^-$ pair ($\ell=e$ or $\mu$), and the $\pi^0$ is reconstructed via the decay  $\pi^0\too\gamma\gamma$. Additionally, the decay $h_c\too\gamma\eta_c$ with $\eta_c\too K^+K^-\pi^0$ is used as the normalization channel.

\section{BESIII detector and Monte Carlo simulation}
The BESIII detector~\cite{besiii} records symmetric $e^+e^-$ collisions
provided by the BEPCII storage ring~\cite{bepcii}, which operates with a peak luminosity of $1\times10^{33}$~cm$^{-2}$s$^{-1}$ in the center-of-mass energy range from 2.0 to 4.95~GeV. BESIII has collected large data samples in this energy region~\cite{whitepaper}. The cylindrical core of the BESIII detector covers 93\% of the full solid angle and consists of a helium-based multilayer drift chamber~(MDC), a plastic scintillator time-of-flight
system~(TOF), and a CsI(Tl) electromagnetic calorimeter~(EMC),
which are all enclosed in a superconducting solenoidal magnet
providing a 1.0~T magnetic field. The solenoid is supported by an
octagonal flux-return yoke with resistive plate counter muon
identification modules interleaved with steel.
The charged-particle momentum resolution at $1~{\rm GeV}/c$ is
$0.5\%$, and the $dE/dx$ resolution is $6\%$ for electrons
from Bhabha scattering. The EMC measures photon energies with a
resolution of $2.5\%$ ($5\%$) at $1$~GeV in the barrel (end cap)
region. The time resolution in the TOF barrel region is 68~ps, while
that in the end cap region is 110~ps. The end cap TOF
system was upgraded in 2015 using multi-gap resistive plate chamber
technology, providing a time resolution of
60~ps~\cite{etof}. About 70\% of the data sample used here was taken after this upgrade.

Simulated data samples produced with a {\sc geant4}-based~\cite{geant4} Monte Carlo (MC) package, which
includes the geometric description of the BESIII detector and the
detector response, are used to determine detection efficiencies
and to estimate backgrounds. The simulation models the beam
energy spread and initial state radiation (ISR) in the $e^+e^-$
annihilations with the generator {\sc kkmc}~\cite{KKMC}.
The inclusive MC sample includes the production of open charm
processes, the ISR production of vector charmonium(-like) states,
and the continuum processes incorporated in {\sc kkmc}~\cite{KKMC}. The known decay modes are modelled with {\sc evtgen}~\cite{ref:evtgen} using branching fractions taken from the
Particle Data Group (PDG)~\cite{pdg}, and the remaining unknown charmonium decays are modelled with {\sc lundcharm}~\cite{ref:lundcharm}. Final state radiation (FSR) from charged final state particles is incorporated using {\sc photos}~\cite{photos}.
Signal MC samples for $\EE \too\pi^+\pi^-h_c$ are generated using isotropic phase space populations at each center-of-mass energy point, assuming that the input line-shape follows a coherent sum of $Y(4220)$ and $Y(4390)$ Breit-Wigner functions, whose parameters are fixed to the measured values in Ref.~\cite{pipihc}. The subsequent $h_c\too\pi^0J/\psi$ decay is generated uniformly in phase space, and the transition $h_c\too\gamma\eta_c$ is generated with an angular distribution of $1+\text{cos}^2\theta^{*}$, where $\theta^{*}$ is the angle of the photon with respect to the $h_c$ helicity direction in the $h_c$ rest frame. Furthermore, the three-body process $\eta_c\too K^+K^-\pi^0$ is simulated according to the measured two-body invariant-mass distributions~\cite{hc2}.

\section{Event selection}
For each charged track, the distance of closest approach to the interaction point is required to be within $\pm10$ cm in the beam direction and within 1 cm in the plane perpendicular to the beam direction. The polar angle ($\theta$) of the charged tracks must be within the fiducial volume of the MDC $(|\!\cos\theta|<0.93)$. Photons are reconstructed from isolated showers in the EMC, which are at least $10^\circ$ away from the nearest charged track. The photon energy is required to be at least 25 MeV in the barrel region $(|\!\cos\theta|<0.80)$ or 50 MeV in the end cap region $(0.86<|\!\cos\theta|<0.92)$. To suppress electronic noise and energy depositions unrelated to the event, the time at which the photon is recorded in the EMC is required to be within $700$ ns of the event start time.

Since the decays $h_c\too\pi^0J/\psi$ and $h_c\too\gamma\eta_c$ result in the final states $\gamma\gamma\pi^+\pi^-\ell^+\ell^-$ and $\gamma\gamma\gamma\pi^+\pi^-K^+K^-$, respectively, candidate events are required to have four charged tracks with zero net charge, at least two photons for $h_c\too\pi^0J/\psi$, and at least three for $h_c\too\gamma\eta_c$. For the $h_c\too\pi^0J/\psi$ decay, tracks with momenta larger than 1.0~GeV/$c$ are assigned to be leptons from $J/\psi$ decay. Otherwise, they are considered as pions. Leptons from the $J/\psi$ decay with an energy deposited in the EMC larger than 1.0~GeV are identified as electrons, and those with less than 0.4~GeV as muons. For the $h_c\too\gamma\eta_c$ decay, information from TOF and $dE/dx$ measurements is combined to form particle identification (PID) likelihoods for the $\pi$, $K$ and $p$ hypotheses. Each track is assigned with a particle type corresponding to the hypothesis of the highest PID likelihood. Exactly two oppositely charged $\pi$ and $K$ each are required in each event. To reduce the background contributions and to improve the mass resolution, a five-constraint kinematic fit is performed for the two channels. The total four-momentum is constrained to the initial four-momentum of the $e^+e^-$ system. Additionally, the invariant mass of the two photons from the $\pi^0$ decay is constrained to the $\pi^0$ mass taken from the PDG~\cite{pdg}. If there is more than one candidate in an event, the one with the smallest $\chi^{2}$ of the kinematic fit is selected. The $\chi^{2}$ is required to be less than 30 for $h_c\too\pi^0J/\psi$ or 40 for $h_c\too\gamma\eta_c$.

To remove backgrounds from $\eta/\omega\too\pi^+\pi^-\pi^0$ in the $h_c\too\pi^0J/\psi$ decay, events with invariant mass $M(\pi^+\pi^-\pi^0)$ in regions around the mass of the $\eta$ or $\omega$, namely $[0.51, 0.58]$~GeV/$c^2$ or $[0.75, 0.81]$~GeV/$c^2$, respectively, are excluded.
The $J/\psi$ signal region is chosen as the range [3.085, 3.115]~GeV/$c^{2}$ in $M(\ell^+\ell^-)$, and sideband regions for studying the non-resonant backgrounds are defined as the ranges $[3.00, 3.06]$ and $[3.14, 3.20]~\text{GeV}/c^{2}$. The $\eta_c$ signal region is chosen to be [2.92, 3.04]~GeV/$c^{2}$ in $M(K^+K^-\pi^0)$, and sideband regions are the ranges $[2.59, 2.71]~\text{GeV}/c^{2}$ and $[3.25, 3.37]~\text{GeV}/c^{2}$.

\section{Branching fraction measurement}
\begin{figure}[tbp]
\centering
\includegraphics[width=.45\textwidth]{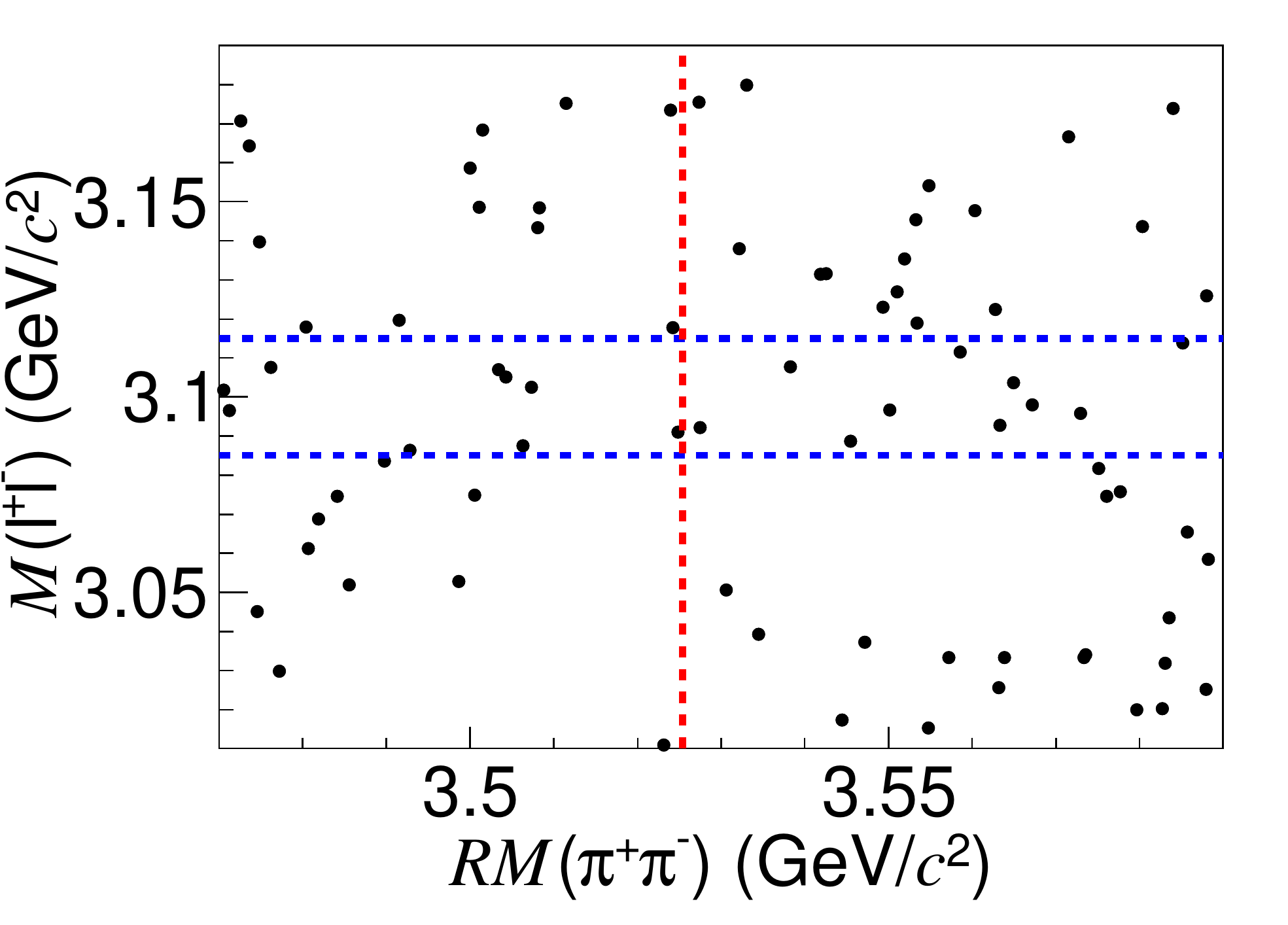}
\includegraphics[width=.45\textwidth]{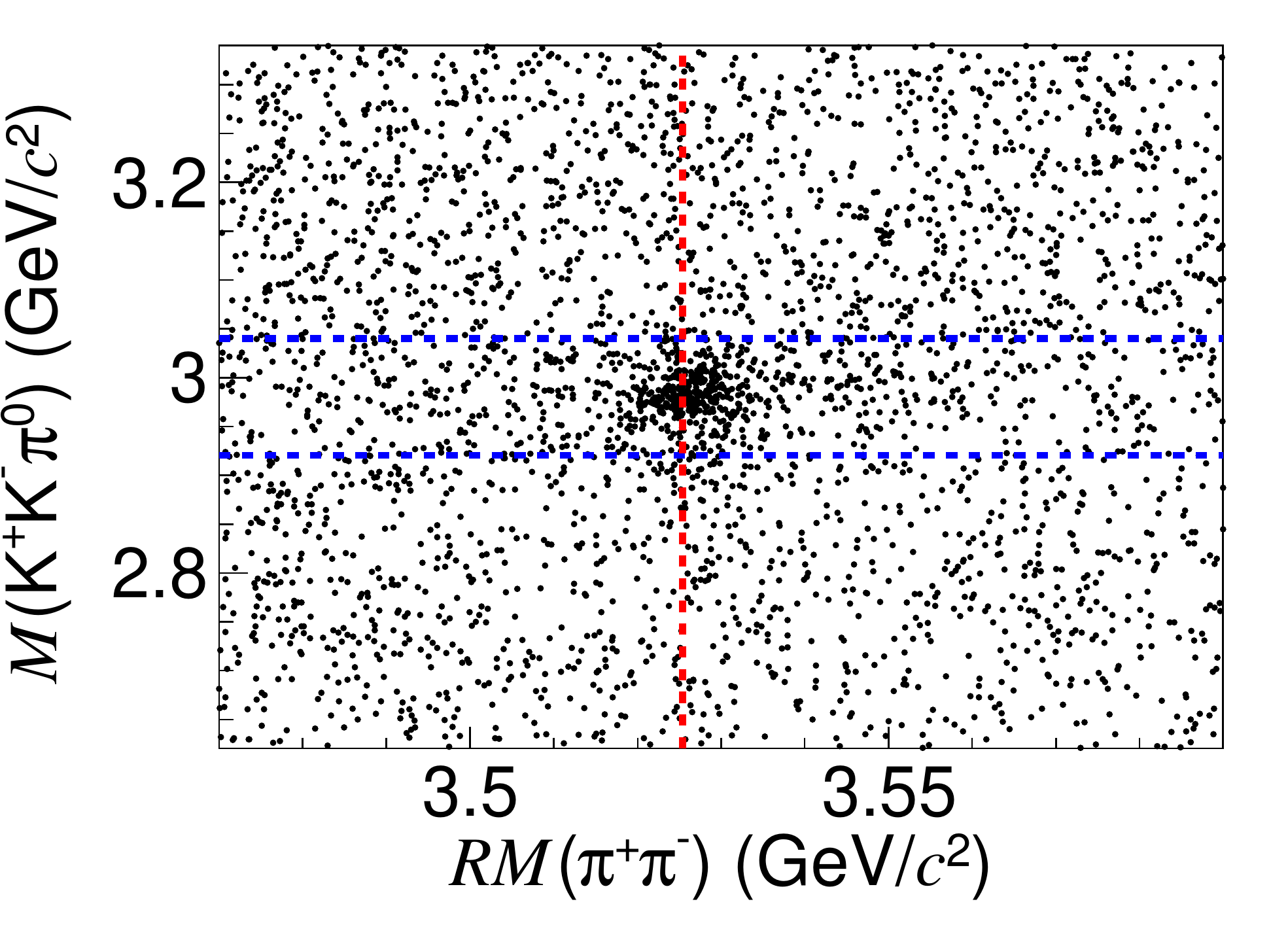}
\caption{\label{fig:scatter} Distributions of $M(\ell^+\ell^-)$ versus $RM(\pi^+\pi^-)$ for $h_c\too\pi^0J/\psi$ (left panel) and $M(K^+K^-\pi^0)$ versus $RM(\pi^+\pi^-)$ for $h_c\too\gamma\eta_c$ (right panel). The horizontal dashed lines denote the signal regions of the $J/\psi$~(left) and $\eta_c$~(right), and the vertical dashed lines mark the nominal $h_c$ mass.}
\end{figure}

Distributions of the invariant mass of $\ell^+\ell^-$ (or $K^+K^-\pi^0$), $M(\ell^+\ell^-)$ (or $M(K^+K^-\pi^0)$), versus the recoil mass of $\pi^+\pi^-$, $RM(\pi^+\pi^-)$ for $h_c\too\pi^0J/\psi$ (or $\gamma\eta_c$) are shown in Fig.~\ref{fig:scatter}. Here, $RM(\pi^+\pi^-)=\sqrt{(P_{e^+e^-}-P_{\pi^+}-P_{\pi^-})^2}$, where $P_{e^+e^-}$ and $P_{\pi^{\pm}}$ are the four-momenta of the initial $e^+e^-$ system and the $\pi^{\pm}$, respectively. No significant signal is observed for the $h_c\too\pi^0J/\psi$ decay. As expected, a high-density area can be observed originating from the $h_c\too\gamma\eta_c$ decay.

\begin{figure}[tbp]
\centering
\includegraphics[width=.45\textwidth]{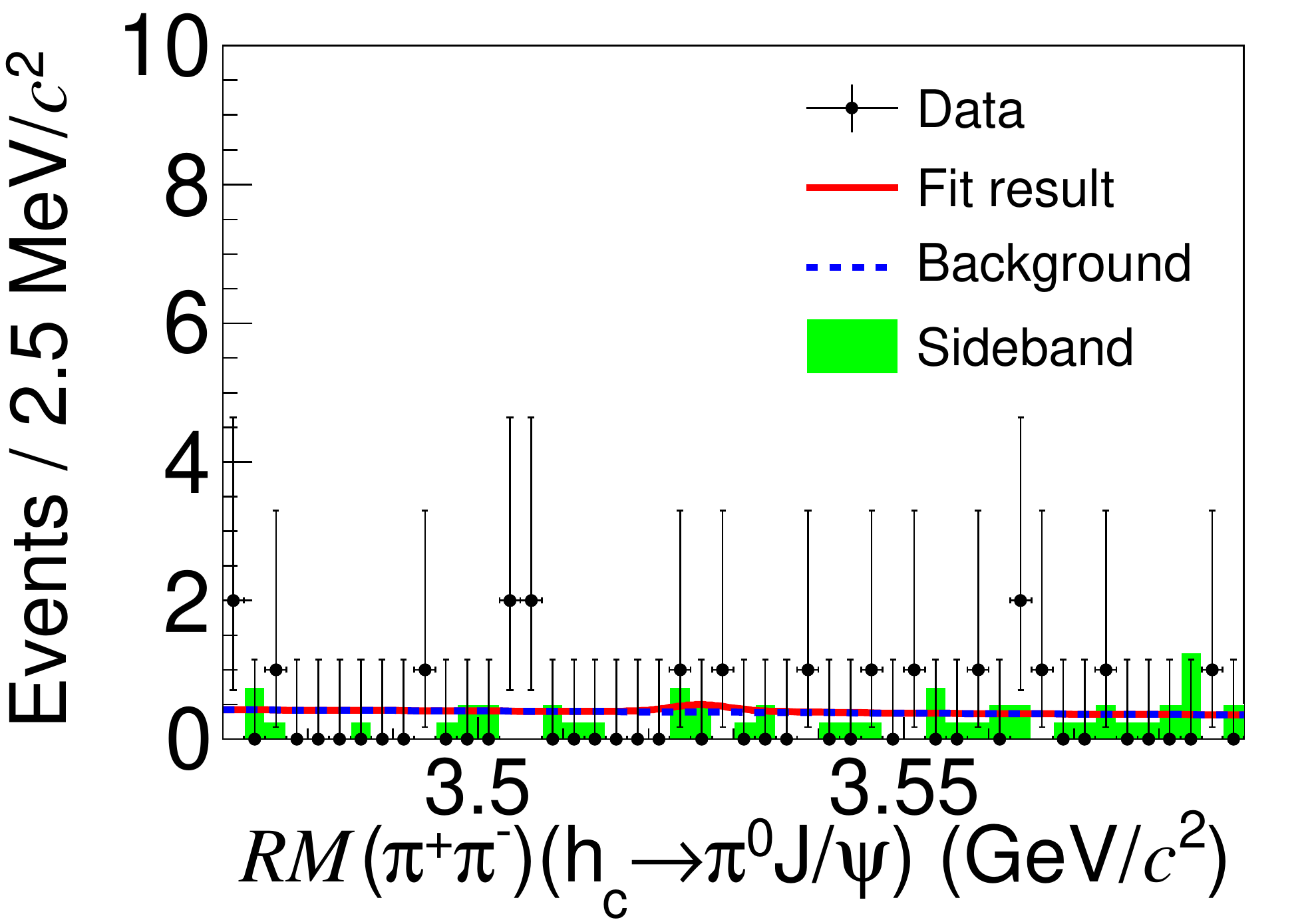}
\includegraphics[width=.45\textwidth]{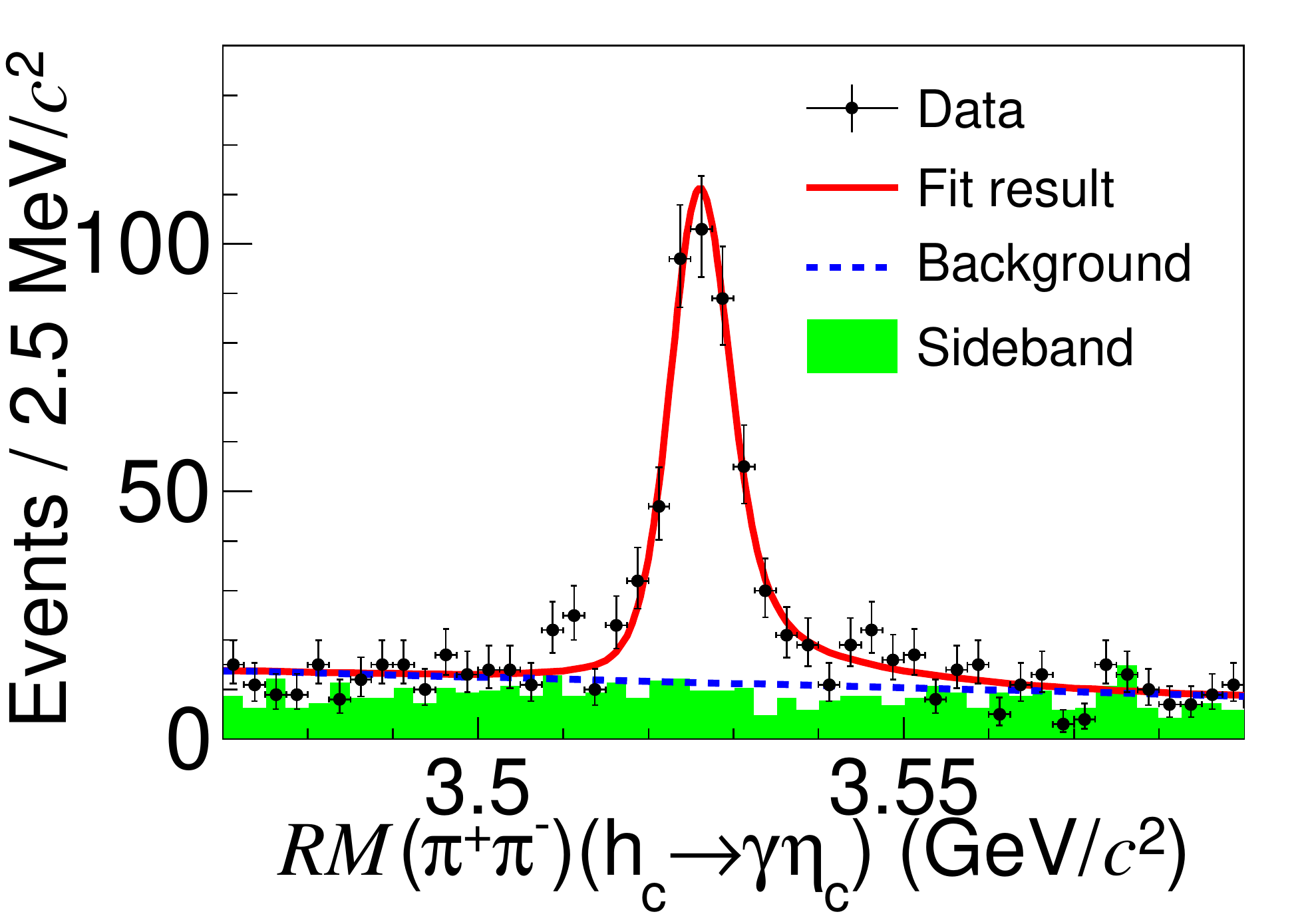}
\caption{\label{fig:hc} Results of the simultaneous fit to the $RM(\pi^+\pi^-)$ distributions from the decays $h_c\too\pi^0J/\psi$ (left panel) and $h_c\too\gamma\eta_c$ (right panel). The red solid lines are the total fit results and the blue dashed lines are the background components. The green shaded histograms correspond to the normalized events from the $J/\psi$ and $\eta_c$ sideband regions.}
\end{figure}

Figure~\ref{fig:hc} shows the distributions of $RM(\pi^+\pi^-)$ for the decays $h_c\too\pi^0J/\psi$ and $h_c\too\gamma\eta_c$ for data in the $J/\psi$ and $\eta_c$ signal regions. No significant signal is seen for the $h_c\too\pi^0J/\psi$ decay, while a clear peak is present for the $h_c\too\gamma\eta_c$ decay. The green shaded histograms correspond to the normalized events from the $J/\psi$ and $\eta_c$ sideband regions. No significant peaks are found in sideband events, and the sideband is not used in the following fit. A detailed study of the inclusive MC sample indicates that there are no peaking background contributions in the $h_c$ signal region~\cite{inclusive}. In order to extract the $h_c$ signal yield, a simultaneous unbinned maximum-likelihood fit to the $RM(\pi^+\pi^-)$ in two decay channels is performed. The signal shape of $RM(\pi^+\pi^-)$ is modeled by a shape obtained from the simulation convolved with a Gaussian function. The mean value and width of the Gaussian function are allowed to float, but are constrained to be the same for the two channels in the simultaneous fit. The background is described by a first-order polynomial function. The solid curves in Fig.~\ref{fig:hc} show the fit results. The $h_c$ signal yields are $0.5\pm1.6$ and $451.6\pm26.7$ for the decays $h_c\too\pi^0J/\psi$ and $h_c\too\gamma\eta_c$, respectively. Since no significant signal is observed for the $h_c\too\pi^0J/\psi$ decay, an upper limit at $90\%$ confidence level (C.L.) using the Bayesian method is given. With the fit function described above, the $h_c\too\pi^0J/\psi$ signal yield is scanned to obtain the likelihood distribution, which is then convolved with the systematic uncertainty. The upper limit on the $h_c\too\pi^0J/\psi$ signal yield $N^{\text{up}}_{\pi^0J/\psi}$ at $90\%$ C.L. is obtained by solving the equation $\int_{0}^{N^{\text{up}}_{\pi^0J/\psi}}F(x)dx/\int_{0}^{\infty}F(x)dx=0.90$, where $x$ is $h_c\too\pi^0J/\psi$ signal yield and $F(x)$ is the probability density function of the likelihood distribution. The upper limit $N^{\text{up}}_{\pi^0J/\psi}$ is determined to be $4.8$.

The branching ratio $\mathcal{B}(h_c\too\pi^0J/\psi)/\mathcal{B}(h_c\too\gamma\eta_c\too\gamma K^+K^-\pi^0)$ is calculated with
\begin{equation}
    \frac{\mathcal{B}(h_c\too\pi^0J/\psi)}{\mathcal{B}(h_c\too\gamma\eta_c\too\gamma K^+K^-\pi^0)} = \frac{N_{\pi^0J/\psi}}{N_{\gamma\eta_c}}
    \frac{\sum\limits_i\mathcal{L}_i\sigma_i(1+\delta_i)\epsilon_i^{\gamma\eta_c}}{\sum\limits_i\mathcal{L}_i\sigma_i(1+\delta_i)\epsilon_i^{\pi^0J/\psi}}
    \frac{1}{\mathcal{B}(J/\psi\too\ell^+\ell^-)},
\end{equation}
where $N$ is the yield of signal events, $\mathcal{L}$ is the integrated luminosity~\cite{luminosity}, $\sigma$ is the cross section of $\EE\too\pi^+\pi^-h_c$~\cite{pipihc}, $1+\delta$ is the radiative correction factor~\cite{KKMC, QED}, $\epsilon$ is the efficiency, $\mathcal{B}(J/\psi\too\ell^+\ell^-)$ is the branching fraction of $J/\psi\too\ell^+\ell^-$ from the PDG~\cite{pdg}, and $i$ denotes each energy point. Table~\ref{tab:sumtable} shows the luminosity, cross section, radiative correction factor and efficiency at each center-of-mass energy. So the ratio $\sum\limits_i\mathcal{L}_i\sigma_i(1+\delta_i)\epsilon_i^{\gamma\eta_c}/\sum\limits_i\mathcal{L}_i\sigma_i(1+\delta_i)\epsilon_i^{\pi^0J/\psi}$ is $0.847$, and the upper limit on the branching ratio $\mathcal{B}(h_c\too\pi^0J/\psi)/\mathcal{B}(h_c\too\gamma\eta_c\too\gamma K^+K^-\pi^0)$ at $90\%$ C.L. is determined to be $\mathcal{B}(h_c\rightarrow\pi^0J/\psi)/\mathcal{B}(h_c\rightarrow\gamma\eta_c\rightarrow\gamma K^+K^-\pi^0)<7.5\times10^{-2}$. With the world averages of $\mathcal{B}(h_c\rightarrow\gamma\eta_c)$ and $\mathcal{B}(\eta_c\rightarrow K^+K^-\pi^0)$ from the PDG~\cite{pdg}, which is $\mathcal{B}(h_c\rightarrow\gamma\eta_c)=(51\pm6)\%$ and $\mathcal{B}(\eta_c\rightarrow K^+K^-\pi^0)=\mathcal{B}(\eta_c\rightarrow KK\pi)/6=(1.217\pm0.067)\%$ according to the isospin symmetry, an upper limit on the branching fraction $\mathcal{B}(h_c\rightarrow\pi^0J/\psi)$ at $90\%$ C.L. can be determined to be $\mathcal{B}(h_c\rightarrow\pi^0J/\psi)<4.7\times10^{-4}$. The systematic uncertainties have been considered in these upper limit calculations.
\begin{table}[tbp]
\centering
\begin{tabular}{c  c  c  c  c  c}
  \hline
  \hline
  \ \ $\sqrt{s}$ (GeV) \ \ & \ \ $\mathcal{L}$ (pb$^{-1}$) \ \ & \ \ $\sigma$ (pb) \ \ & \ \ $1+\delta$ \ \ & \ \ $\epsilon^{\pi^{0}J/\psi}$ $(\%)$ \ \ & \ \ $\epsilon^{\gamma \eta_c}$ $(\%)$ \ \ \\
  \hline
  4.189 & 570 & 15.315 & 0.705 & 17.60 & 15.07 \\
  4.199 & 526 & 25.183 & 0.706 & 17.68 & 15.25 \\
  4.209 & 572 & 38.706 & 0.712 & 17.58 & 15.33 \\
  4.219 & 570 & 51.740 & 0.725 & 17.59 & 15.27 \\
  4.226 & 1101& 57.072 & 0.740 & 18.41 & 15.62 \\
  4.236 & 530 & 57.956 & 0.766 & 17.80 & 15.38 \\
  4.244 & 594 & 55.185 & 0.786 & 17.62 & 15.40 \\
  4.258 & 828 & 49.053 & 0.817 & 17.58 & 15.04 \\
  4.267 & 531 & 46.002 & 0.831 & 16.94 & 15.03 \\
  4.278 & 176 & 43.657 & 0.842 & 16.85 & 14.47 \\
  4.288 & 502 & 42.757 & 0.847 & 16.38 & 14.10 \\
  4.312 & 546 & 44.609 & 0.845 & 16.79 & 14.36 \\
  4.338 & 505 & 51.729 & 0.834 & 17.29 & 14.45 \\
  4.358 & 544 & 58.972 & 0.829 & 17.74 & 14.83 \\
  4.378 & 579 & 63.700 & 0.835 & 17.99 & 14.63 \\
  4.397 & 508 & 61.392 & 0.857 & 17.42 & 14.42 \\
  4.416 & 1091& 52.183 & 0.898 & 17.12 & 14.03 \\
  4.437 & 570 & 39.299 & 0.960 & 15.85 & 13.34 \\
  \hline
  \hline
\end{tabular}
\caption{\label{tab:sumtable} The luminosity $\mathcal{L}$, cross section $\sigma$, radiative correction factor $1+\delta$ and efficiency $\epsilon$ at each center-of-mass energy $\sqrt{s}$.}
\end{table}

\section{Systematic uncertainty}
The sources of systematic uncertainty related to the branching ratio $\mathcal{B}(h_c\too\pi^0J/\psi)/\mathcal{B}(h_c\too\gamma\eta_c\too\gamma K^+K^-\pi^0)$ and branching fraction $\mathcal{B}(h_c\too\pi^0J/\psi)$ are summarized in Table~\ref{tab:sumerror}, where the uncertainties associated with the charged track PID efficiencies and normalization channel yields, as well as the branching fraction of the normalization channel used to obtain the absolute branching fraction of $h_c\too\pi^0J/\psi$.

The uncertainty in the photon efficiency and charged track PID efficiency is $1\%$ per photon or per track~\cite{track1, track2, track3}. The uncertainty due to the kinematic fit is estimated by correcting the helix parameters of charged tracks, and the difference between the results with and without this correction is taken as the uncertainty~\cite{helix}. To estimate the uncertainty related to the input line-shape of the process $\EE\too\pi^+\pi^-h_c$, the input line-shape is changed to a flat line-shape and the difference is taken as the uncertainty.

The process $\EE\too\pi^+\pi^-h_c$ is generated by a three-body phase-space model. The uncertainty related to the MC decay model is obtained by substitution with a model $\EE\too\pi^{\pm}Z_c(4020)^{\mp}\too\pi^+\pi^-h_c$. The angular distribution of $h_c\rightarrow\pi^0J/\psi$ is varied from phase space to $1\pm\text{cos}^2\theta^{**}$, where $\theta^{**}$ is the angle of the $\pi^0$ with respect to the $h_c$ helicity direction in the $h_c$ rest frame. These two items are combined as the total uncertainty from the MC decay model.

The uncertainty from the $J/\psi$ mass window requirement is estimated using $\EE \too \gamma_{ISR}\psi(3686)\too\gamma_{ISR}\pi^{+}\pi^{-}J/\psi$ events~\cite{jpsimasswindow}. The efficiency difference between data and MC simulation is taken as the systematic uncertainty, which arises from the different mass resolutions in the data and the simulation. The uncertainty from the $\eta_c$ mass window requirement is obtained by shifting the $\eta_c$ mass window by $\pm10~\text{MeV}/c^{2}$, and we take the difference of $\mathcal{B}(h_c\rightarrow\gamma\eta_c\rightarrow\gamma K^+K^-\pi^0)$ to the nominal one as the systematic uncertainty. These two items are combined as the total uncertainty of the mass window.

The number of events for the normalization channel $h_c\too\gamma\eta_c$ is determined to be $N_{\gamma\eta_c}=451.6\pm26.7$. The uncertainty from the yield of the normalization channel is thus $5.9\%$. The uncertainties from the branching fractions $\mathcal{B}(J/\psi\too l^+l^-)$, $\mathcal{B}(h_c\too \gamma\eta_c)$ and $\mathcal{B}(\eta_c\too K^+K^-\pi^0)$ are taken from the PDG~\cite{pdg}.

The overall systematic uncertainties except for the fit procedure are obtained by adding all sources of systematic uncertainties in quadrature, assuming they are uncorrelated, and are summarized in Table~\ref{tab:sumerror}.
\begin{table}[tbp]
\centering
\begin{tabular}{c c c}
  \hline
  \hline
  Source & $\mathcal{B}(h_c\too\pi^0J/\psi)/\mathcal{B}(h_c\too\gamma\eta_c\too\gamma K^+K^-\pi^0)$ & $\mathcal{B}(h_c\too\pi^0J/\psi)$  \\
  \hline
  Photon efficiency               & 1.0 & 1.0  \\
  PID efficiency                  & 4.0 & 4.0  \\
  Kinematic fit                   & 2.0 & 2.0  \\
  Input line-shape                & 1.2 & 1.2  \\
  MC decay model                  & 1.6 & 1.6  \\
  Mass window                     & 2.4 & 2.4  \\
  Normalization channel yield     & 5.9 & 5.9  \\
  $\mathcal{B}(J/\psi\too l^+l^-)$& 0.6 & 0.6  \\
  $\mathcal{B}(h_c\too \gamma\eta_c)$   & $-$ & 11.8  \\
  $\mathcal{B}(\eta_c\too K^+K^-\pi^0)$ & $-$ &  5.5  \\
  \hline
  Sum                             & 8.1 & 15.3  \\
  \hline
  \hline
\end{tabular}
\caption{\label{tab:sumerror} Relative systematic uncertainties (in $\%$) on the branching ratio $\mathcal{B}(h_c\too\pi^0J/\psi)/\mathcal{B}(h_c\too\gamma\eta_c\too\gamma K^+K^-\pi^0)$ and the branching fraction $\mathcal{B}(h_c\rightarrow\pi^0J/\psi)$. Dashes are used when sources of uncertainty are not applicable.}
\end{table}

The sources of uncertainty in the fit procedure include the fit range and background shape. The limit of the fit range is varied by $\pm$5 MeV/$c^{2}$, and the background shape is replaced from the first-order polynomial to a constant function or a second-order polynomial function. For the uncertainty from the fit procedure for the upper limit measurement, different combinations of fit range and background shape are used to get the upper limits, and the largest upper limit is chosen as the nominal one. Then the likelihood distribution of the nominal upper limit is convolved with the above overall systematic uncertainties to get the final upper limit.

\section{Summary and discussion}
The decay $h_c\rightarrow\pi^0J/\psi$ is searched for using the process $e^+e^-\rightarrow\pi^+\pi^-h_c$
with data samples collected at center-of-mass energies between 4.189 and $4.437~\text{GeV}$ with the BESIII detector corresponding to a total integrated luminosity of 11 fb$^{-1}$.
No significant signal is observed for the decay channel $h_c\rightarrow\pi^0J/\psi$. The upper limits on the branching ratio $\mathcal{B}(h_c\too\pi^0J/\psi)/\mathcal{B}(h_c\too\gamma\eta_c\too\gamma K^+K^-\pi^0)$ and the branching fraction $\mathcal{B}(h_c\rightarrow\pi^0J/\psi)$ at $90\%$ confidence level are determined to be $7.5\times10^{-2}$ and $4.7\times10^{-4}$, respectively. The latter is derived from the former using the measured branching fraction of the normalization channel. This is the first upper limit measurement of the branching fraction for the decay $h_c\rightarrow\pi^0J/\psi$. The measured results are not consistent with the measurements by the E760 Collaboration~\cite{e760}, while in agreement with the E835 Collaboration~\cite{e835}. The $h_c$ total width is $0.7\pm0.4$ MeV from the PDG~\cite{pdg}, and we take 1.1 MeV as the $h_c$ total width conservatively. Therefore, the upper limit on the partial width for $h_c\rightarrow\pi^0J/\psi$ is conservatively determined to be $\Gamma(h_c\too\pi^0J/\psi)<0.52$ keV, which is one order-of-magnitude lower than the current theoretical predictions (several keV)~\cite{theory1, theory2, theory3, theory4}.

\acknowledgments

The BESIII collaboration thanks the staff of BEPCII and the IHEP computing center for their strong support. This work is supported in part by National Key R\&D Program of China under Contracts Nos. 2020YFA0406300, 2020YFA0406400; National Natural Science Foundation of China (NSFC) under Contracts Nos. 11905179, 11625523, 11635010, 11735014, 11822506, 11835012, 11935015, 11935016, 11935018, 11961141012, 12022510, 12025502, 12035009, 12035013, 12061131003; the Chinese Academy of Sciences (CAS) Large-Scale Scientific Facility Program; Joint Large-Scale Scientific Facility Funds of the NSFC and CAS under Contracts Nos. U1732263, U1832207; CAS Key Research Program of Frontier Sciences under Contract No. QYZDJ-SSW-SLH040; 100 Talents Program of CAS; INPAC and Shanghai Key Laboratory for Particle Physics and Cosmology; ERC under Contract No. 758462; Nanhu Scholars Program for Young Scholars of Xinyang Normal University; European Union Horizon 2020 research and innovation programme under Contract No. Marie Sklodowska-Curie grant agreement No 894790; German Research Foundation DFG under Contracts Nos. 443159800, Collaborative Research Center CRC 1044, FOR 2359, FOR 2359, GRK 214; Istituto Nazionale di Fisica Nucleare, Italy; Ministry of Development of Turkey under Contract No. DPT2006K-120470; National Science and Technology fund; Olle Engkvist Foundation under Contract No. 200-0605; STFC (United Kingdom); The Knut and Alice Wallenberg Foundation (Sweden) under Contract No. 2016.0157; The Royal Society, UK under Contracts Nos. DH140054, DH160214; The Swedish Research Council; U. S. Department of Energy under Contracts Nos. DE-FG02-05ER41374, DE-SC-0012069.

\clearpage

\section*{The BESIII collaboration}
\addcontentsline{toc}{section}{The BESIII collaboration}
\begin{small}
M.~Ablikim$^{1}$, M.~N.~Achasov$^{10,b}$, P.~Adlarson$^{68}$, S. ~Ahmed$^{14}$, M.~Albrecht$^{4}$, R.~Aliberti$^{28}$, A.~Amoroso$^{67A,67C}$, M.~R.~An$^{32}$, Q.~An$^{64,50}$, X.~H.~Bai$^{58}$
% ...
\\
\\
{\it
$^{1}$ Institute of High Energy Physics, Beijing 100049, People's Republic of China\\
% ...
$^{4}$ Bochum Ruhr-University, D-44780 Bochum, Germany\\
% ...
$^{10}$ G.I. Budker Institute of Nuclear Physics SB RAS (BINP), Novosibirsk 630090, Russia\\
% ...
$^{14}$ Helmholtz Institute Mainz, Staudinger Weg 18, D-55099 Mainz, Germany\\
% ...
$^{28}$ Johannes Gutenberg University of Mainz, Johann-Joachim-Becher-Weg 45, D-55099 Mainz, Germany\\
% ...
$^{32}$ Liaoning Normal University, Dalian 116029, People's Republic of China\\
% ...
$^{50}$ State Key Laboratory of Particle Detection and Electronics, Beijing 100049, Hefei 230026, People's Republic of China\\
% ...
$^{58}$ University of Jinan, Jinan 250022, People's Republic of China\\
% ...
$^{64}$ University of Science and Technology of China, Hefei 230026, People's Republic of China\\
% ...
$^{67}$ University of Turin and INFN, (A)University of Turin, I-10125, Turin, Italy; (B)University of Eastern Piedmont, I-15121, Alessandria, Italy; (C)INFN, I-10125, Turin, Italy\\
$^{68}$ Uppsala University, Box 516, SE-75120 Uppsala, Sweden\\
% ...
$^{b}$ Also at the Novosibirsk State University, Novosibirsk, 630090, Russia\\
% ...
}
\end{small}

\end{document}